\documentclass[preprint]{aastex}

\newcommand{\flux}{ergs cm$^{-2}$ s$^{-1}$}

\newcommand{\lumi}{ergs s$^{-1}$}
\newcommand{\col}{cm$^{-2}$}

\title{
Discovery of a new pulsating X-ray source with a 1549.1-s period, AX J183220$-$0840
}

\author{Mutsumi Sugizaki}
\affil{Tsukuba Space Center, National Space Development Agency of Japan}
\affil{2-1-1 Sengen, Tsukuba, Ibaraki, 305-8505, Japan}
\email{sugizaki.mutsumi@nasda.go.jp}
\author{Kenzo Kinugasa}
\affil{Gunma Astronomical Observatory}
\affil{6860-86 Nakayama, Takayama, Agatsuma, Gunma 377-0702, Japan}
\author{Keiichi Matsuzaki}
\affil{The Institute of Space and Astronautical Science}
\affil{3-1-1 Yoshinodai, Sagamihara, Kanagawa, 229-8510, Japan}
\author{Yukikatsu Terada}
\affil{Department of Physics, University of Tokyo}
\affil{Bunkyo-ku, Tokyo 113-0033, Japan}
\author{Shigeo Yamauchi}
\affil{Faculty of Humanities and Social Sciences, Iwate University}
\affil{3-18-34 Ueda, Morioka, Iwate, 020-8550, Japan}
\and
\author{Jun Yokogawa}
\affil{Department of Physics, Kyoto University}
\affil{Kyoto 606-8502, Japan}

\begin{abstract}

A new pulsating X-ray source, AX J183220$-$0840, with a 1549.1-s period
was discovered at R.A.$= 18^{\rm h} 32^{\rm m} 20^{\rm s}$ and Dec.$=-8^{\circ}40'30''$
(J2000, uncertainty$=0\farcm 6$) during an ASCA observation on the Galactic plane.
The source was observed two times, in 1997 and in 1999.
A phase-averaged X-ray flux of $1.1\times 10^{-11}$ {\flux} 
and pulsation period of $1549.1\pm 0.4$ s 
were consistently obtained from these two observations.
The X-ray spectrum was represented by a flat absorbed power-law 
with a photon-index of $\Gamma\simeq 0.8$ and 
an absorption column density of $N_{\rm H}\simeq 1.3\times10^{22}$ {\col}.
Also, a signature of iron K-shell line emission with a centroid of 6.7 keV
and an equivalent width of approximately 450 eV was detected.
From the pulsation period and the iron-line feature,
AX J183220$-$0840 is likely to be a magnetic white dwarf binary
with a complexly absorbed thermal spectrum with a temperature of about 10 keV.

\end{abstract}


\begin{document}

\maketitle

\section{Introduction}

Since many X-ray sources in the Galaxy are considered to be compact objects, 
the population of X-ray sources 
gives us information on the evolution and endpoints 
of stars.
Thus, the study of faint X-ray sources in the Galaxy
is an important subject in X-ray astronomy.
However, because of the difficulty in hard X-ray optics,
imaging observations with high sensitivities have been 
limited to the soft X-ray band below 3 keV.
Since the soft X-rays with an energy below 3 keV are 
substantially absorbed by the interstellar medium (ISM) in the Galaxy,
our knowledge of X-ray sources has always been limited to 
that of the local area neighboring our solar system.
Recently, this situation has begun changing 
with the discoveries of faint X-ray pulsars 
by deep observations of ASCA and SAX
(e.g., Sugizaki et al. 1997; Kinugasa et al. 1998; Torii et al. 1999; 
Oosterbroek et al. 1999; Sakano \& Koyama 2000).

ASCA is the first X-ray astronomical satellite which allows imaging observations
with a high sensitivity in the X-ray band above 3 keV \cite{Tanaka1994}.
The ASCA Galactic plane survey, 
aimed at the systematic study of Galactic X-ray sources,
was performed from March 1997 to April 1999.
During the observation in the direction of $(l,b)=(23 \fdg 0, 0 \fdg 0)$ in the Galactic plane survey,
a new variable X-ray source, AX J183220$-$0840, 
was detected in the GIS field of view        
\cite{Sugizaki1999}.
As a result of analyzing source variabilities, 
a signature of a coherent pulsation with a 1550-s period was discovered from the light curve.
Since the exposure time of 6 ks in the survey observation 
was too short to investigate the variabilities with the 1550-s period
and the SIS data could not be utilized for a quantitative analysis
because of the bad source position,
we conducted a follow-up observation and 
succeeded in confirming the periodicity and obtaining detailed spectral properties.

In this paper, we report the discovery of a new X-ray pulsation source, AX J183220$-$0840, 
and the results of analyzing data 
obtained from the Galactic plane survey and the follow-up observation.
We discuss the identification of the source based on these results.

\section{Observation and analysis}

The data processing and analysis below
were performed according to the standard
analysis procedure supported by the ASCA Guest Observer Facility at NASA/GSFC.

AX J183220$-$0840 was first observed with ASCA on 1997 October 11 
during an observation in the direction of $(l,b)=(23\fdg 0, 0\fdg 0)$
in the ASCA Galactic plane survey.
The observation was conducted in the PH-nominal mode for the GIS 
and in 4CCD-faint/bright mode for the SIS
with an 6-ks exposure time after data screening. 
The source was detected in the GIS field of view, 20$'$-apart from the optical axis.
Since the source was located at a corner of the SIS field of view,
the SIS data in this observation could not be utilized for a quantitative analysis.
We denote this first observation as OBS\#1.

We performed a follow-up observation of AX J183220$-$0840 on 1999 October 17--18,
and denote this observation as OBS\#2.
The source was observed at the 1CCD nominal position
in the PH-nominal mode for the GIS 
and in the 1CCD-faint/bright mode for the SIS.
The exposure times were 34 ks for the GIS and 28 ks for the SIS after data screening.

Table \ref{tab:obslog} summarizes the logs of the two observations, OBS\#1 and OBS\#2.

\subsection{Source position}

Since the spatial resolution of the GIS is inferior to that of the SIS,
we redetermined the position of AX J183220$-$0840 using the SIS image of OBS\#2.
The refined position is 
(R.A., Dec.)$=(18^{\rm h} 32^{\rm m} 20^{\rm s}, -8^{\circ}40'30'')$(J2000)
or $(l, b)=(23\fdg 0404, 0\fdg 2576)$, where the position determination accuracy 
is $0\farcm 6$ at the 90\% confidence limit.
The positions determined from the two observations, OBS\#1 and OBS\#2,
are in complete agreement within the positional accuracy.

We investigated known X-ray catalog sources and found that
the position is in accord with WGA J1832.3$-$0840 in the ROSAT WGA source catalog
\cite{White1994}
within the position accuracy.
However, no other information about this source has been revealed so far.

\subsection{Light curves and timing analysis}

Due to the unfavorable source position in the SIS during OBS\#1
and because the exposure time of the SIS was shorter than that of the GIS 
after data screening,
we used the GIS data for timing analysis.
Events of the sources were extracted from a circle of 6$'$ radius for the GIS,
and the arrival times were corrected to the solar system barycenter.

\begin{figure}[p]
\begin{center}
\includegraphics[angle=-90, width=16cm]{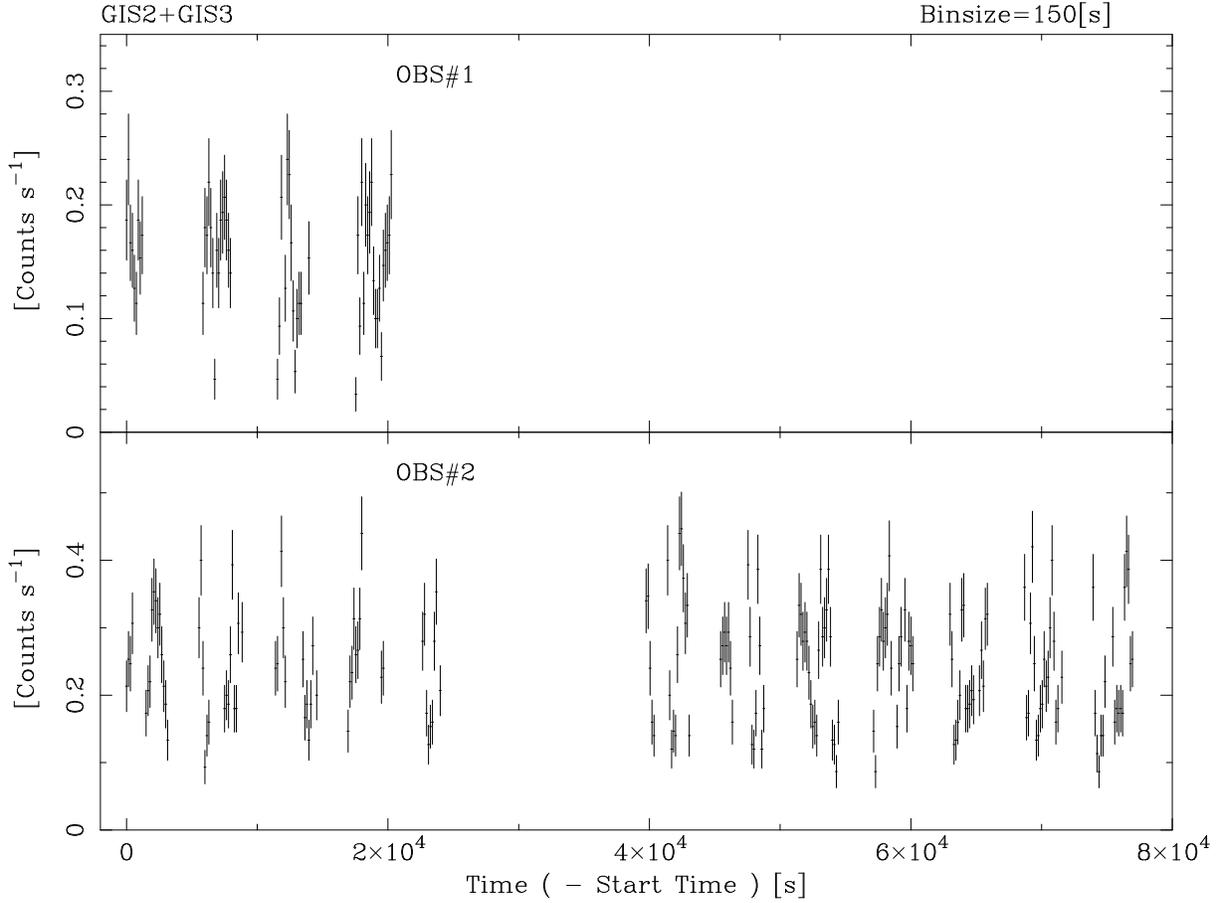}
\caption{
Light curves of the GIS2$+$GIS3 in OBS\#1 and in OBS\#2.
Start times of OBS\#1 and OBS\#2 are described in Table \ref{tab:obslog}.
The difference in the average count rate between the two observations
is caused by the position dependence of the effective area.
Taking account of that effect,
the average source fluxes in OBS\#1 and in OBS\#2 are consistent.
}
\label{fig:lc}
\end{center}
\end{figure}

Figure \ref{fig:lc} shows the light curves of the GIS
in OBS\#1 and in OBS\#2.
Large variabilities are apparent in both the curves.
The average source fluxes in OBS\#1 and in OBS\#2
were consistent,
taking account of the position dependence of the effective area
in the GIS field of view.

\begin{figure}[p]
\begin{center}
\includegraphics[angle=-90, width=15cm]{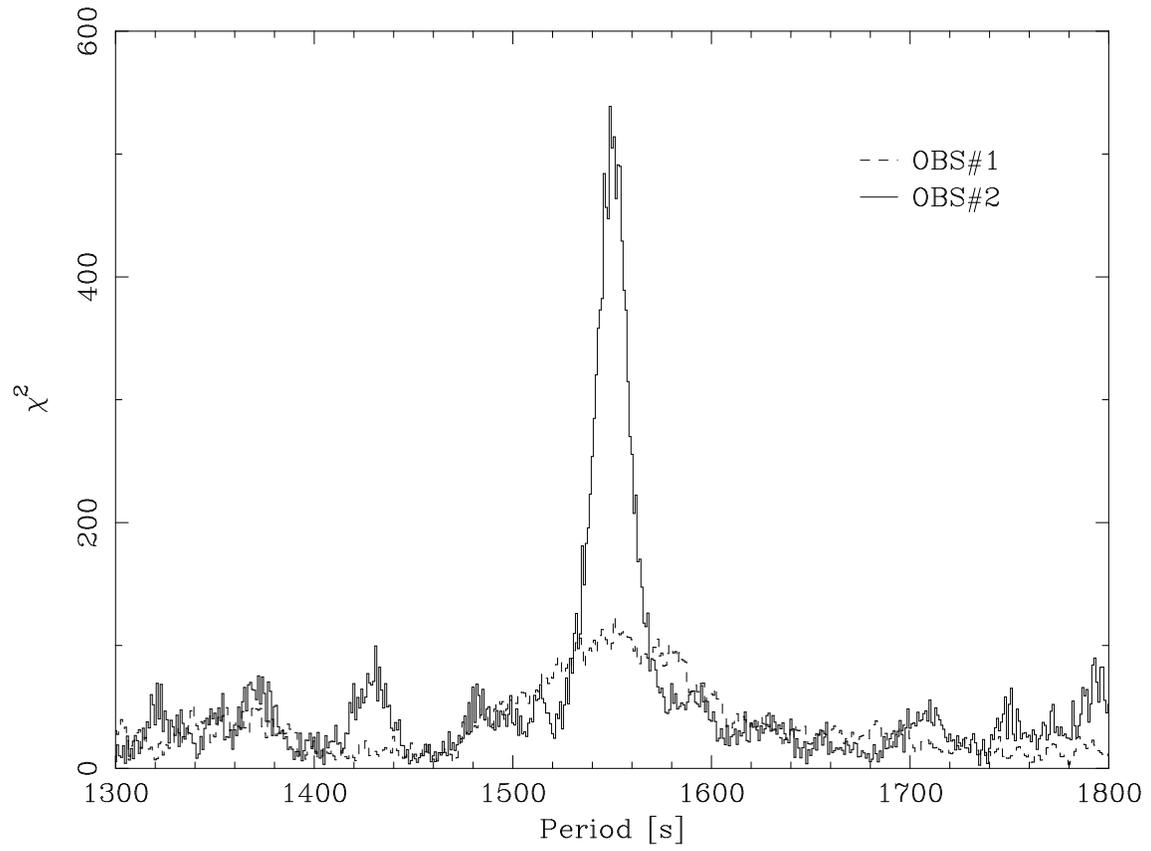}
\caption{
Periodograms obtained from the GIS2$+$GIS3 light curves in OBS\#1 and OBS\#2.
}
\label{fig:efs}
\end{center}
\end{figure}

We performed epoch-folding analysis
to investigate the periodicity of the variation.
Figure \ref{fig:efs} shows the periodogram obtained from the light curves
of the GIS in OBS\#1 and OBS\#2.  
The periodicity is clearly seen from a large significant peak at 1549 s.
We performed a pulse arrival-time analysis (e.g., Sugizaki et al. 1997)
by dividing the OBS\#2 observation interval
into three segments 
and determined the best period to be $1549.1 \pm 0.4$ s 
with a 90\% confidence limit. 
Figure \ref{fig:eflc} shows the folded light curves of the GIS
with the best-fit period of 1549.1 s 
in energy bands of 0.7--2.5 keV, 2.5--5.0 keV, 5.0--10 keV, and 0.7--10 keV
obtained from the OBS\#2 data. 
All the pulse profiles are represented by a double-peak profile and
there is no large difference among these energy bands.
The pulsed fraction in the total (0.7--10 keV) band is 63\%,
corrected for background.
We searched for a change of the pulsation period 
between observations OBS\#1 and OBS\#2, and 
during each observation interval.
However, we found no significant change of the period.
The upper limit of the period derivative was determined to 
be $|\dot{P}|<8\times 10^{-5}$ s s$^{-1}$ with a 99\% confidence limit
from pulse arrival-time analysis in OBS\#2.

We investigated the long-term variation of the flux averaged over pulse phase.
However, no significant variation with an amplitude exceeding 14\% in RMS (root-mean-square)
was detected at the 99\% confidence limit.
The PSPC counting rate of WGA J1832.3$-$084
in the ROSAT WGA catalog, 0.0165 cnts s$^{-1}$ \cite{White1994},
was also consistent with the soft X-ray flux measured 
in the ASCA observations.

\subsection{Spectral Analysis}

We performed a spectral analysis mainly using the SIS data in OBS\#2.
Since the energy resolution of the SIS is superior to that of the GIS,
the SIS has good sensitivity for line emissions.
Also, a combined analysis of OBS\#1 and OBS\#2 is not preferred 
since the accuracy of the response function for the source in OBS\#1 
is substantially degraded
because of the large off-axis angle ($\simeq 20'$) of the source position in OBS\#1.
We used the other data to check the consistency and confirmed the results.
The source spectra were collected from a 3$'$-radius circle on the SIS detectors,
and the background spectra were collected from the source-free region 
in the same field of view of each detector.

\begin{figure}[p]
\begin{center}
\includegraphics[angle=-90, width=15cm]{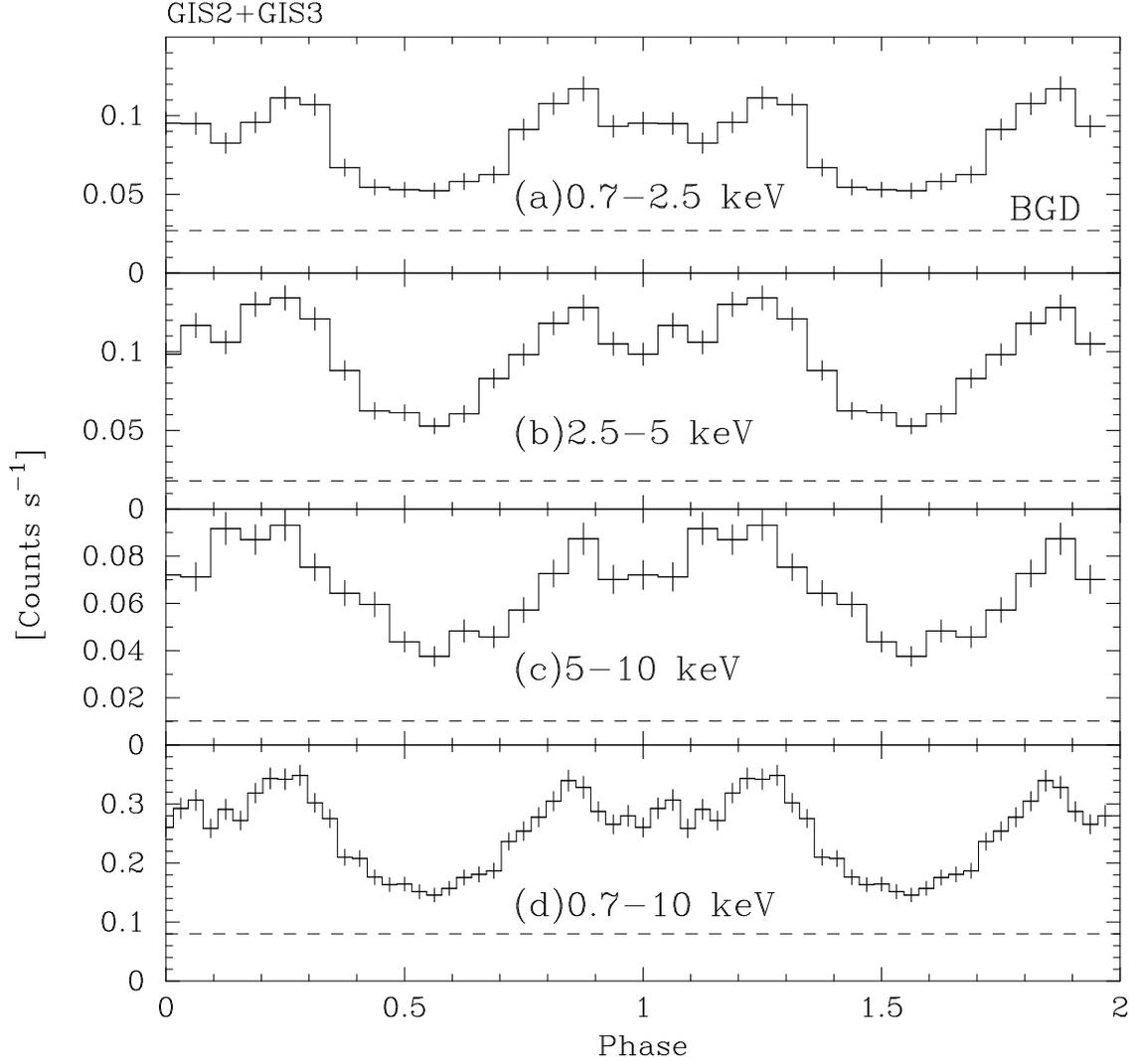}
\caption{
Folded light curves with a best-fit pulsation period of 1549.1 s 
in the (a) 0.7--2.5 keV, (b) 2.5--5.0 keV, (c) 5.0--10 keV, and (d) 0.7--10 keV bands
obtained from the GIS data in OBS\#2.
Dashed lines indicate the non-source background levels.
}
\label{fig:eflc}
\end{center}
\end{figure}

We firstly defined a phase of 0.34--0.66 as the minimum and
a phase of 0.75--1.31 as the maximum in Figure \ref{fig:eflc}
and extracted phase-sorted energy spectra.
However, we saw no difference in spectral shapes 
between the minimum and the maximum phases 
within the statistical accuracy.
This is in agreement with the result that 
the pulse profiles are consistent among the different energy bands
from 0.7 to 10 keV (see Figure \ref{fig:eflc}).

\begin{figure}[p]
\begin{center}
\includegraphics[angle=-90, width=15cm]{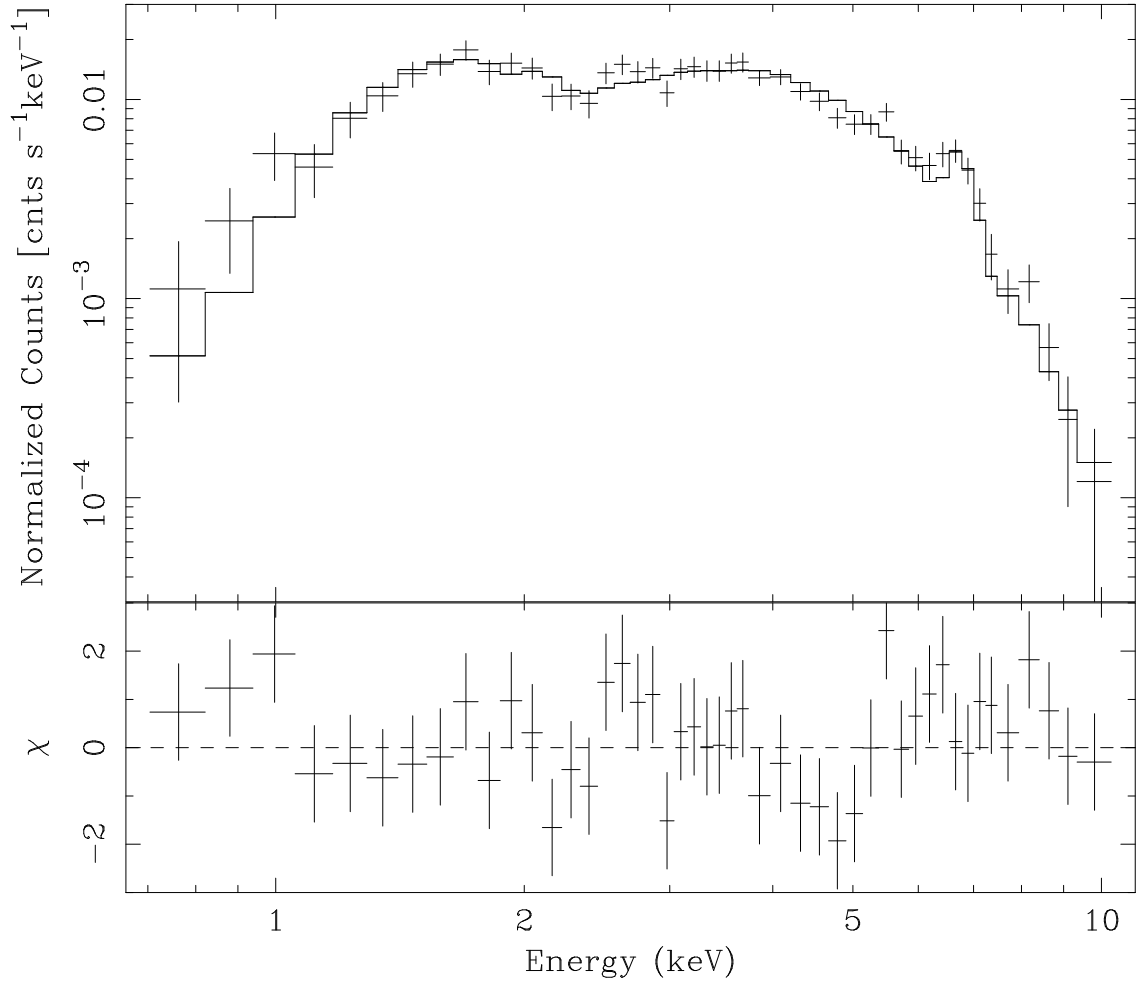}
\caption{
Phase-averaged X-ray spectra obtained by the SIS in OBS\#2.
The solid line is the best-fit, multi-absorption, thin-thermal 
emission model (see text).
The lower panel shows the residuals from the best-fit model.
}
\label{fig:spectra}
\end{center}
\end{figure}

We thus analyzed the phase-averaged energy spectra accumulating 
all live-time data in OBS\#2.
Figure \ref{fig:spectra} shows an X-ray spectrum taken by the SIS;
the spectrum represents a very hard spectral shape.
We fitted the spectrum with a power-law model 
with an absorption of the ISM.
The reduced chi-squared ($\chi^2_\nu$) of 1.65 is derived for 44 degrees of freedom ($\nu$),
which is not acceptable within the 99\% confidence limit.
Since the residuals show a signature of iron K-shell emission lines,
we added a narrow Gaussian line to the power-law model and fitted it to the data. 
The $\chi^2_\nu$ is reduced to 1.21 for 42 degrees of freedom, 
which is acceptable within the 90\% confidence limit.
The significance of the iron line component is estimated to be greater than 99\%
from the $F$ test.
We obtained a power-law index of $0.76_{-0.15}^{+0.15}$, 
an absorption column density of $1.27_{-0.23}^{+0.26}$ {\col},
and an iron line centroid of $E_{\rm Fe}=6.73_{-0.13}^{+0.11}$ keV
from the best-fit model,
where the errors represent the 90\% confidence limits (hereafter). 
The unabsorbed flux in the 0.7--10 keV band 
and the equivalent width of the iron line emission 
are estimated to be $1.1\times 10^{-11}$ {\flux} and 450$^{+160}_{-160}$ eV 
from the best-fit model.
To examine the width of the iron line, 
we floated the width of the Gaussian function and performed the fitting again.
However, the improvement was not significant. 

The obtained line center energy ($E_{\rm Fe}\simeq 6.7$ keV) and 
the equivalent width ($EW\simeq 450$ eV) of the iron K-line 
are in accord with the properties of a thin-thermal plasma emission 
with a temperature of about 10 keV.
We next examined a model of the thin-thermal plasma emission.
However, the flat continuum spectrum represented by the power-law with index $\Gamma\simeq 0.7$
cannot be represented by a simple, one-component thermal emission model.
However, if we introduce a multi-component thermal spectrum 
with different absorption column densities,
we can reproduce such a flat spectrum, 
which is observed in many magnetic white dwarf binaries \cite{Ezuka1999}.
Thus, we fitted the data with a thermal-equilibrium plasma emission model   
\citep{Raymond1977}
with interstellar absorption plus partial covering fraction absorption.
We assumed a solar abundance for the plasma.
The fit is acceptable within the 90\% confidence limit ($\chi^2_\nu=1.18$; $\nu=42$).
We obtained a plasma temperature of $10.6_{-2.1}^{+3.6}$ keV,
a partial covering absorption column density of $10.6_{-2.1}^{+3.6}\times 10^{22}$ {\col},
and a partial covering fraction of $0.65_{-0.09}^{+0.08}$
from the best-fit model.
Also, we estimated the unabsorbed flux in the 0.7--10 keV band 
to be $1.5\times 10^{-11}$ {\flux}. 
The best-fit model of the multi-component thermal spectrum
is shown together with the data in Figure \ref{fig:spectra}.

All the results of the spectral fitting are summarized in Table \ref{tab:spec}.

\section{Discussion}

We now discuss the nature of the new X-ray pulsating source, AX J183220$-$0840,
based on the results of the analysis.

We first estimate the distance to the source from the X-ray absorption column density,
which has a correlation with the hydrogen column density
within the RMS $\lesssim$ 50\% \cite{Ryter1975,Savage1977}. 
We obtain the absorption column density for the source, 
$N_{\rm H} \simeq 1.3\times 10^{22}$ {\col},
from the spectral analysis.
The hydrogen column density of the Galactic ISM in the direction to the source 
can be estimated from the H{\sc i} and H$_2$ densities. 
The H{\sc i} density is derived from the radio 21-cm observation \cite{Dickey1990}, 
which is $N_{\rm H_I}=1.9\times 10^{22}$ {\col}.
The H$_2$ density is estimated from the CO-line intensity \cite{Dame1987}
and CO-to-H$_2$ conversion factor \cite{Hunter1997}, which is 
$N_{\rm H_2}=2.1\times 10^{22}$ {\col}.
Thus, the hydrogen column density in the direction of the source 
accumulated over the Galaxy
is evaluated to be $N_{\rm H_I} + 2N_{\rm H_2}=6.1\times 10^{22}$ {\col}.
Assuming that ISM is uniformly distributed on the Galaxy plane 
within the 10 kpc from the Galactic center
and the solar system is 8 kpc from the Galactic center,
the absorption column density of $N_{\rm H} \simeq 1.3\times 10^{22}$ {\col}
corresponds to a distance of $4$ kpc.
Thus, the luminosity of the source is estimated to 
be $L_{\rm X} = 2.0\times 10^{34} d_{\rm 4kpc}^2$ {\lumi}
from the unabsorbed source flux, $1.1\times 10^{-11}$ \flux. 

The 1549.1-s pulsation period of AX J183220$-$0840 
is in accord with typical rotation periods of white dwarfs (WDs)
observed in cataclysmic variables (CVs).
The broad, double-peak pulse profile and 
the X-ray spectrum represented by a multi-absorption, thin-thermal plasma emission
with a temperature of about 10 keV also agree with 
those of typical magnetic CVs (e.g., Mukai et al. 1994; Ishida et al. 1997; Ezuka \& Ishida 1999).
The estimated luminosity of $2.0\times 10^{34} d_{\rm 4kpc}^2$ {\lumi} seems to be relatively high
compared with typical values of the magnetic CVs.
However, it can be reasonably explained if the absorption is partially intrinsic to the source
and the distance is less than the 4 kpc estimated by assuming that 
the absorption is responsible for the ISM.
Thus, a magnetic WD binary is considered to be the most plausible candidate
for the X-ray origin of AX J183220$-$0840.

If AX J183220$-$0840 is a binary neutron star (NS) pulsar, 
the pulsation period is as long as that of the slowest known X-ray pulsar, 
RX J0146.9$+$6121, which has a 1407.3-s period \cite{Habel1998}.
The hard X-ray spectrum represented by a power-law with an index of $\Gamma\simeq 0.8$ 
agrees with that of typical NS binaries \cite{Nagase1989}.
However, the iron line feature with a centroid of $6.7$ keV
indicating that it originates from He-like iron ion, 
disagrees with the typical NS binaries, many of which have a 6.4 keV line 
from neutral iron. 
This argues against the hypothesis of a NS pulsar.

Recently, the number of faint X-ray pulsars with long periods exceeding 400 s
has increased by the discoveries of ROSAT, ASCA, and SAX 
\cite{Israel1998,Torii1999,Oosterbroek1999,Sakano2000}.
It is not clear yet whether some of these are NS binaries or WD binaries.
No evidence of binary systems such as 
eclipses, modulations of the pulsation period, flux changes,
or binary counter parts, has been obtained so far 
from these sources including AX J183220$-$0840.
Since the typical X-ray luminosity of the WD binaries 
($L_{\rm X}\sim 10^{30-32}$ {\lumi}; e.g., Ezuka \& Ishida 1999),
is less by five orders of magnitude than that of the NS binaries
($L_{\rm X}\sim 10^{35.5-37.5}$ {\lumi}; e.g., Nagase 1989),
the number of the WD binaries 
was only about a sixth of that of the NS binaries 
among bright Galactic X-ray sources 
with a flux above $2\times 10^{-11}$ {\flux}
\cite{Bradt1983}.
However, the population of the WD binaries 
is expected to be greater by three orders of magnitude than 
that of the NS binaries in the Galaxy \cite{Patterson1984}.
Thus, AX J183220$-$0840 is considered to be a candidate
of faint WD binaries that 
have been undetectable due to the large distance from the solar system
but have now become detectable by the sensitive observations
of ROSAT and ASCA.

Further X-ray observations are required to confirm the 
identification of AX J183220$-$0840.
Optical observations would also be useful to search for a counterpart,
and possibly to measure binary parameters.

\acknowledgments

We thank all the members 
of the ASCA Galactic plane survey team.
We are also grateful to M. Cripe
for his careful review of the manuscript.



\begin{table}[p]
\begin{center}
\caption{Summary of observation log}\label{tab:obslog}
\begin{tabular}{cccccc}
\hline
\hline
Observation     & Date  & Start time & End time & \multicolumn{2}{c}{Exposure time}\\
       & (UT)  & (MJD)      & (MJD)    & (ks/GIS) & (ks/SIS)\\
\hline 
OBS\#1 & 1997 Oct 11     & 50732.11030 & 50732.33845 & 6.2  & --- \\
OBS\#2 & 1999 Oct 17--18 & 51468.73552 & 51469.65668 & 34.2 & 27.9 \\  
\hline
\end{tabular}
\end{center}
\end{table}

\begin{table}[p]
\begin{center}
\caption{Summary of spectral fitting results}\label{tab:spec}
\begin{tabular}{lcccc}
\hline
\hline
                      & \multicolumn{4}{c}{Model}\\
Parameter             & PL       & PL+Gaus & PL+Gaus & Pcabs$*$RS \\
\hline
$\Gamma$              & $0.66_{-0.14}^{+0.14}$ & $0.76_{-0.15}^{+0.15}$ & $0.81_{-0.16}^{+0.16}$ & ---  \\	
$kT$ [keV]            & --- & --- & --- & $10.6_{-2.2}^{+3.6}$  \\	
$N_{\rm H}$ [$10^{22}${\col}] & $1.17_{-0.22}^{+0.25}$ & $1.27_{-0.23}^{+0.26}$ & $1.33_{-0.25}^{+0.28}$ & $1.41_{-0.33}^{+0.32}$ \\	
$N_{\rm H,pc}^{\rm (1)}$ [$10^{22}${\col}] & --- & --- & --- & $10.6_{-2.1}^{+3.6}$ \\	
$f_{\rm pc}^{\rm (2)}$            & --- & --- & --- & $0.65_{-0.09}^{+0.08}$ \\	
$E_{\rm Fe}$ [keV]       & --- & $6.73_{-0.13}^{+0.11}$ & $6.74_{-0.12}^{+0.12}$ & --- \\	
$\sigma_{\rm Fe}$ [keV] & ---   & 0:frozen & $0.22_{-0.16}^{+0.15}$ & --- \\	
{\it EW}$_{\rm Fe}^{\rm (3)}$ [eV] & --- & 450$^{+160}_{-160}$ & 670$^{+270}_{-240}$ & --- \\	
$F_{\rm 0.7-10}^{\rm (4)}$ [10$^{-11}${\flux}] & 1.1 & 1.1 & 1.1 & 1.5 \\	
\hline 
$\chi^2_\nu/\nu$              & 1.65/44  & 1.21/42 & 1.13/41 & 1.18/42 \\	
\hline
\end{tabular}
\tablenotetext{}{
All errors represent the 90\% confidence limits. 
}
\tablenotetext{(1)}{Partial covering absorption column density.}
\tablenotetext{(3)}{Partial covering fraction.}
\tablenotetext{(3)}{Equivalent width of iron line.}
\tablenotetext{(4)}{Flux in the 0.7--10 keV band corrected for the effect of absorption.}
\end{center}
\end{table}


\end{document}